\documentclass[12pt]{article}
\usepackage{astrobib,graphicx}

\newcommand{\gtapprox}{\raisebox{-0.5ex}{$\,\stackrel{>}{\scriptstyle
\sim}\,$}}
\newcommand{\ltapprox}{\raisebox{-0.5ex}{$\,\stackrel{<}{\scriptstyle
\sim}\,$}}

\begin{document}

\title{Filaments in the Galactic Center -- with special reference to the ``Snake''}

\author{Geoffrey V. Bicknell, $^{1,2}$ 
Jianke Li $^{1,3}$
} 

\date{}
\maketitle

{\center
$^1$ Research School of Astronomy \& Astrophysics, Australian National University. Mt Stromlo
Observatory, Cotter Road, Weston, ACT, Australia, 2611 \\Geoff.Bicknell@anu.edu.au\\[3mm]
$^2$ Department of Physics \& Theoretical Physics, Australian National University, Canberra, ACT,
Australia, 0200 \\[3mm]
$^3$ Department of Mathematics, Australian National University, Canberra, ACT, Australia, 0200\footnote{Current
address: Higher Education, Department of Education, Training and Youth Affairs.} \\[3mm] }

\begin{center}
\em Accepted for publication in the Proceedings of the Astronomical Society of Australia.
\end{center}
\begin{abstract}
The nonthermal filaments in the Galactic
Center constitute one of the great mysteries of this region of the Galaxy. We summarise the observational
data on these filaments and critically review the various theories which currently outnumber the observed
filaments. We summarise out theory for the longest of these filaments, the Snake, and discuss the relevance of
this model for the other filaments in the Galactic Center region. The physics involved in our model for the Snake
involves much of the physics that has dominated the career of Professor Don Melrose. In particular, the diffusion
of relativistic electrons in the Snake is determined from the theory of resonant scattering by Alfv\'en waves.  
\end{abstract}


\section{Introduction}

We would like to thank the organisers of this Festschrift for Professor Don Melrose for the opportunity
to contribute this paper, summarising our recent research \cite{bicknell01c} on one of the mysterious
filaments in the Galactic Center, known as the Snake. First, however, on this occasion, we would like
to record a few personal notes based on our somewhat different interactions with Don Melrose. Geoff first met
Don, when he took up his first appointment at Mt. Stromlo and Siding Spring Observatories (now the Research
School of Astronomy and Astrophysics) at the Australian National University. Geoff had started working on
extragalactic jets at that time and this clearly involved the physics of particle acceleration on which Don
was an expert and he was not. When Geoff started at Mt. Stromlo, he contacted Don, and the result of the
ensuing collaboration was a paper
\cite{bicknell82a}, that created a lot of interest and half of which he would still defend today. Which
half? The half that clearly sets out the relationship between hydrodynamic turbulent input and its
dissipation via particle acceleration at the high wave number end of a turbulent cascade, to which they
both contributed. That collaboration with an outstanding scientist was one of the most memorable
experiences of his career at that time and one that he always looks back upon with satisfaction. Jianke
worked in the ARC Special Research Centre for Theoretical Astrophysics that Don headed from 1991 to
2000. Jianke was impressed with Don's vision for creating a career path for young theoretical
astrophysicists and with his capacity for research and administration. Jianke also wrote a paper with
Don on pulsar magnetospheres \cite{li94a} and during the course of this work was impressed by Don's
strongly focussed approach to Science, by his physical insight and by his insistence on understanding a
problem fully at every stage. We both wish Don well in continuing his wide interests in Astrophysics,
and in his new and challenging position as head of the School of Physics at Sydney University. We trust
that he will continue to contribute to Astrophysics well into the future.

The work that is the subject of this paper fortuitously involves physics to which Don has made
significant and enduring contributions. Our work on the curious filament in the
Galactic Center known as ``The Snake'' involves the following: 
\begin{itemize}
\item Magnetic fields.
\item Particle acceleration.
\item Resonant scattering of relativistic particles and the diffusion of relativistic particles in the
interstellar medium.
\item A strong connection with solar physics.
\end{itemize} 
Those of you who are familiar with Don's work will know that he has contributed significantly to all of these
areas. The observations of the Snake also involved the outstanding Ph.D. thesis work of Sydney University
student Andrew Gray and his advisors and colleagues Professor Lawrence Cram, Professor Ron Ekers, Dr. Miller
Goss and Dr. Jenny Nicholls. The Snake was discovered through observations with the Molonglo Synthesis
Telescope \cite{gray91a}. It is therefore doubly appropriate to discuss this work on this occasion.

In this paper, as well as summarising the work in our recent ApJ Letter, we have taken the opportunity to 
review some of the recent theoretical papers on the filaments, to  expand some of the details in our own
recent paper, to extend the treatment of relative timescales by incorporating a discussion of the important
constraint imposed by synchrotron cooling,  and to indicate where the physics discussed in this paper may
relate to the Galactic Center filaments as a whole. 

\section{Filaments in the Galactic Center}

The Snake is one of a number of filaments in the central $\sim 100 \> \rm pc$ of the Galaxy that are regarded
as one of the ``major mysteries'' of the Galactic Center. The first set  of such filaments, and by far the
most visually spectacular, the ``Arc'', were discovered by \citeN{yusef84a}. Since that seminal discovery a
number of other filamentary systems have been discovered -- many with quite prosaic names such as the
``Threads''
\cite{morris85a,lang99b}, the ``Snake'' \cite{gray91a,gray95a} and the ``Pelican'' \cite{anantha99a,lang99c}. (For
an impressive panoramic view of the Galactic Center at a wavelength of 90 cm, showing a number of these
filaments, see the image in the paper by
\shortciteN{larosa00a}.) The Snake was discovered in a survey by the Molonglo Synthesis Telescope
 and has been the subject a recent paper by us \cite{bicknell01c}. However, an account of our
model for the Snake would be incomplete without a discussion of the context in which this work was done.
Therefore, in this section we give a brief review of the observational and theoretical work that has been carried
out on these curious filaments. In the following section we separately address one of the outstanding
issues raised by the observations - the strength of the magnetic field in the inner 100 parsecs
of the Galaxy. For a comprehensive review of the  observational and theoretical situation up until 1996 see
the review by \citeN{morris96a}.

\subsection{Observations}

The main observational characteristics of the filaments observed in the central regions of the Galaxy are:
\begin{enumerate}
\item They are narrow. Generally the filaments are about a fraction of a parsec in transverse extent. The Arc is
wider than this $\sim 5 \> \rm pc$ but breaks up into a number of subfilaments each a fraction of a parsec
in width and separated by about 0.5~pc \cite{yusef87a}. 
\item The emission is nonthermal, highly linearly polarised and the magnetic field is aligned along the
filaments.
\item There are substantial rotation measure variations associated with the environments of the filaments.
\item There is a wide variety in spectral index characteristics. The Arc has an inverted spectral index,
$\alpha \approx -0.3$ ($F_\nu \propto \nu^{-\alpha}$). However, this steepens considerably away from the galactic
plane where the Arc degenerates into a more diffuse structure -- the northern and southern plumes
\cite{reich90a,pohl92a}. Most of the various isolated non-thermal filaments usually have a steeper spectral
index,
$0.6 < \alpha < 0.4$ at frequencies below 1.4~GHz that is consistent with what is usually observed for
optically thin non-thermal spectra. The spectral index is steeper, $\alpha > 1.5$, at frequencies higher
than 5~GHz. If the break in spectral index at $\sim 5 \> \rm GHz$ can be attributed to synchrotron
cooling in a $\sim \rm mG$ field, then the implied age $\sim 2 \times 10^4 \> \rm  yr$. The Snake is singular
in that the spectral index has a conventional value, $\alpha \approx 0.4 - 0.5$ near the kinks in its
structure but this flattens with distance away from the major kink.
\item When imaged at high resolution, many filaments have a multi-stranded appearance. \citeN{yusef97a} found
that the filamentary system G359.54+0.18 is double and that the polarised emission is greatest where the
two filaments appear to cross each other. \citeN{lang99b} found that both the northern and southern Threads
have bifurcated regions and the northern thread becomes diffuse at its northwestern extremity. The Pelican
(G358.85+0.47) consists of multiple parallel strands which also become more evident near its ends
\cite{lang99c}. The Snake shows a bifurcation at its major kink \shortcite{gray95a}. In some cases the
filaments to be braided. e.g. the filaments associated with the HII region, Sagittarius~C \cite{liszt95a}.
\item Almost all of the filaments, except the Pelican, are aligned close to perpendicular to the Galactic
plane. This has been taken by some to indicate a large scale poloidal $\sim \rm mG$ field in the Galactic
Center region and \shortcite{lang99c} have suggested that the Pelican may mark a transition region of
magnetic orientation some 225~pc in projected distance from the centre of the Galaxy. 
\end{enumerate}

A feature that these filaments have in common is that they all seem to involve magnetic fields that are
strong compared to what we are used to in the interstellar medium. This has led to suggestions of a large
scale, high strength ($\sim \rm mG$) magnetic field in the Galactic Center region (e.g. \citeN{morris98a}).
On the other hand, \citeN{roberts99a} has pointed out that the field strengths are of order that estimated
in the cores of molecular clouds and that observation is more consistent with the theory for the Snake that
we outline below. The issue of the strength of the magnetic field in the Galactic Center region is 
discussed in \S~\ref{s:b-field}.

\subsection{Theory}

There have been numerous widely different physical theories advanced for the filaments in the Galactic
Center. Indeed the number of theories exceeds the number of filaments, surely representing a triumph of the
inventiveness and creativity of the human spirit! Such a situation is also characteristic of a subject in its
infancy. A number of theories for the Galactic filaments were reviewed by
\shortciteN{gray95a}. Here we concentrate on some of the more recent attempts but also include some of the
older models that have survived to the present or which have played a significant role in our attempts to
understand these filaments. 

The theories form naturally into a number of classes:
\begin{enumerate}
\item Star-based models incorporating some form of interaction interaction of a star or cluster of stars with
the interstellar medium.
\item Interaction of a hypothetical galactic wind with molecular clouds.
\item Interaction of moving molecular clouds with a large scale $\sim \> \rm mG$ magnetic field.
\item Electrodynamic models:
\begin{enumerate} 
\item Conventional MHD models involving the transport of magnetic field in the flux freezing
approximation together with reconnection at specific locations.
\item Unconventional models focusing on the currents produced by electric fields interacting with clouds and
the subsequent generation of magnetic fields and instabilities.
\end{enumerate}
(Classifying electrodynamic models in this way is not meant to imply any {\em a priori} value judgement on the
relative merits of such models.)
\item Shock waves in the interstellar medium.
\item Morphologically unusual supernova remnants.
\item Exotic models. An example is the proposal that cosmic strings are responsible for the filaments.
\end{enumerate}

In all but the last class, magnetic fields are implicated. In some of the magnetic field models, the field is
initially $\sim 10 \mu \rm G$ and is amplified by some process or other; in others, a large $\sim \rm mG$
magnetic field pervading the Galactic Center region is assumed. 

Following \citeN{rosner96a}, we list the following criteria that a successful model of the filaments has
to satisfy:
\begin{enumerate}
\item{\bf Energetics.} There needs to be a physical mechanism for the acceleration of relativistic
electrons and this has to account for the luminosity of each filament.
\item{\bf Spectrum.} As we have seen the spectra of the filaments is quite varied. A good model should account
for this variation as well as the individual cases.
\item{\bf Geometry.} A mechanism for the generation of filamentary structures is required. The variation in
structure (straight in some cases, braided in others, kinked in one case) requires explanation. All but one of
the filaments is perpendicular to the galactic plane. 
\item{\bf Location.} A good model has to account for the numerous filaments in the Galactic Center region and 
their apparent association with star formation regions and molecular clouds -- together with the absence of
filaments associated with such regions outside of the Galactic Center.
\end{enumerate}

\subsubsection{Star-based models.} 
\citeN{nicholls95a} proposed a model for the Snake in which a star trail
caused by a rapid runaway star opened up a conduit in the ISM. This model was discussed fully by
\shortciteN{gray95a}.

\citeN{rosner96a} proposed that the source of relativistic electrons is the terminal bow shock of the wind
from a massive star, or cluster of stars. The electrons are transported away from the shock in the direction
of the local interstellar magnetic field. This creates a flux tube whose lateral extent $\sim \rm pc$ is
determined by the size of the stellar wind bubble and which is loaded by relativistic electrons
from the wind shock. The synchrotron cooling from relativistic electrons generates a cooling instability and
this causes the flux tube to contract laterally generating a magnetic field about 30 times larger. 

This model does not seem to have much observational support. For example there is no evidence for massive
stars at any of the radio-bright regions of the various filaments, although such stars are difficult to
detect in the near IR because of their blue colours and the lack of strong spectral features (McGregor,
private communication). Perhaps more importantly, the existence of a synchrotron cooling instability is
problematical: The pressure in a conventional nonthermal plasma with electron spectral index greater than
2,  is dominated by the lowest energy particles and the cooling is dominated by the highest energy particles
so that cooling does not greatly affect the pressure, rendering collapse through cooling difficult. A
let-out here is that the flat spectral index of the synchrotron emission may indicate a flat electron
distribution in which the pressure, as well as the radiation, is dominated by the highest energies. However,
the spectral index along many filaments is a function of position and this is not taken into account in the
model.

\subsubsection{Interaction of a galactic wind with molecular clouds.} 
\citeN{shore99a} invoked the interaction
of a magnetised galactic wind with molecular clouds as the origin of the filaments. In their theory the
filaments are analogous to cometary tails. The magnetic field in the wind ``wraps around'' the molecular cloud
forming a current sheet in its wake. The lateral size of the filament is of order the size of the molecular
cloud and the field is amplified by stretching in the wake until it reaches equilibrium with the ram pressure
of the wind. Stochastic turbulence generated by the unstable current sheet is invoked as the mechanism for
particle acceleration. In their theory, the ultimate source of energy is the the total magnetic energy in the
wake. However, there is no physical description as to how this source of energy is coupled to the turbulence.
Normally, in a turbulent cascade, the dissipated power is related to the energy in large scale eddies. For
many of the filaments, one can make a case for the general physical morphology proposed by Shore and La~Rosa.
In the \citeN{yusef84a} arcs, for example, there are molecular clouds associated with radio-bright
regions. However, in the Snake, there are no molecular clouds at the bright spots that are coincident with the
kinks. The general direction of filaments approximately perpendicular to the Galactic plane is consistent with
this model. However, the discovery of two system of filaments parallel to the galactic plane
\cite{lang99c,larosa01a} seems to contradict it.

\subsubsection{Cloud -- magnetic field interactions}

Serabyn and Morris and their colleagues have been responsible for much of the detailed observational
work of the nonthermal and molecular regions in the vicinity of the Galactic Center. In the course of this
observational program 
\citeN{serabyn94a} have proposed an interesting explanation for the first set of Galactic Center filaments
observed -- the Arc. Their model involves the interaction of a 
fast-moving molecular cloud with an ambient mG magnetic field. The leading face of the cloud is ionised by
radiation from the nearby HII region and the turbulent interaction of the face with the mG interstellar
magnetic field leads to reconnection and acceleration of monoenergetic relativistic electrons. These stream
away from the face of the cloud at the Alfv\'en speed. The estimate of $B \sim \rm mG$ stems from the
equilibration of turbulent and magnetic pressure in the clouds. They argue that this magnetic field must
be characteristic of the entire region and not localised to the cloud for otherwise the filaments would
expand laterally outside of the molecular clouds which do not fill the arc region but instead are clumped
inside it. Whether one accepts this part of the theory or not, and we discuss this further below in
\S~\ref{s:b-field}, Serabyn
\& Morris present good arguments that the relativistic electrons are the consequence of the
cloud-magnetic field interaction since the onset of radio emission in each filament is at the edge of the
molecular clouds. As they point out, this could indicate either emission or absorption by the molecular
structure. However, they suggest that the increase in disorder of the nonthermal filaments away from the
clouds leading to plume-like structures at some distance from the galactic plane and the steepening of the
spectral index with distance from the galactic plane both indicate that the nonthermal particles originate
near the molecular/ISM interface. These plumes may be a key indicator of the dynamics of the Arc and
surrounding structures and are clearly evident in the \shortciteN{larosa00a} image.

The spectrum of these particular nonthermal filaments described by $F_\nu \propto \nu^{0.3}$, suggested to
Serabyn \& Morris a monoenergetic electron spectrum with the electrons accelerated by electrostatic fields
in the reconnection region. However, such a spectral index is also characteristic of synchrotron emission
from an arbitrary distribution, at frequencies lower than that corresponding to the low energy cutoff. In
either case, the monoenergetic or minimum Lorentz factor $\gtapprox 500$, corresponds approximately to
$0.25 \>
\rm GeV$. Whether the spectrum is monoenergetic or something more complicated, Serabyn \& Morris note that
acceleration to such energies using Petschek reconnection would occur over about $10^{14} cm \sim 0.8 \> \rm
mas$ given that particles stream out of the reconnection zone at the Alfv\'en speed. Thus, VLBI observations
could possibly resolve the reconnection region given enough sensitivity.

\citeN{pohl92a} have also modelled the radio emission from the southern plume that seems to be
connected to the Arc. The observational data indicate a decreasing spectral index along the plume and they
have modelled this reasonably successfully using a steady state model which incorporates the injection of
monoenergetic electrons, their cooling due to synchrotron and inverse Compton emission and their diffusion
due to scattering. The calculated  synchrotron emission takes account of the widening flux tube. This model
has much in common with the model that we have proposed for the Snake, except that theirs is steady state
model and ours is time-dependent, both of these approaches being appropriate to the given circumstances. An
interesting point of consistency is their value for the diffusion coefficient $\sim (2.4 -10) \times 10^{25}
\> \rm cm^2 \> s^{-1}$ (see \S~\ref{s:timescales}). It would be interesting to revisit the
\shortciteN{pohl92a} model and investigate the effect of more general electron spectra.

\subsubsection{Electrodynamic models}

\paragraph{Conventional MHD models.} \citeN{heyvaerts88a} attempted to explain the \shortciteN{yusef84a} Arc
by postulating the ejection of coronal loops at speeds $\sim 1000 \> \rm km \> s^{-1}$ from the black hole at
the Galactic Center. These were then supposed to interact with dense gas in the vicinity of Sagittarius~A
producing the Arc via reconnection processes. This idea was criticised by \citeN{morris89a}
on the basis of fine-tuning: Compared to the travel time of the loops from the black hole, the interaction
time is relatively small. It was also criticised by \citeN{benford88a} on the basis of requiring a special
viewing angle and there being no comparable interaction on the opposite side of the Galactic Center.

Our own theory for the Snake \cite{bicknell01c}, also falls into the class of conventional MHD models. This
is discussed in more detail below.

\paragraph{Unconventional models.} In conventional MHD one calculates the electric current from the curl of
the magnetic field whose evolution is either described by the flux-freezing approximation or a description
that incorporates diffusive processes, eg. reconnection. Benford, on the other hand, motivated by his
research in laboratory plasmas, has focused on the current in a theory for the Arc and the Snake
\citeN{benford88a,benford97a}. In his theory, an electric field, ${\bf E} = c^{-1} ({\bf v}_c \times {\bf
B}) $ is produced by the interaction of a conducting cloud of velocity ${\bf v}_c$ with the pre-existing
magnetic field, ${\bf B}$ whose magnitude is of order a mG. In the boundary layer at the edge of the cloud,
the current and electric field are perpendicular to the magnetic field but then link up with the magnetic
field outside the cloud. A current circuit is formed with the current parallel to the field outside
the cloud and, in addition, a return current is established in the ISM. In the case of the Arc he suggests
that the current may return from the plume-like ends of the Arc and find its way through the ISM in the
vicinity of the Galactic Center or may link up with other filaments south of the Galactic plane. The
resistance of the circuit, resulting from scattering of the electrons by ion acoustic turbulence generated in
the current loop, is a key element of the theory. In his theory for the Arc, the Joule dissipation due to
this resistance is the energy source for particle acceleration and the resultant radio emission.

The current generates a toroidal field. In the case of the Arc, Benford estimates a small radius $\sim
10^9 \> \rm cm$  per current filament, requiring congregation of filaments to form discernible
structures and predicting that the observed filaments should break up into smaller
subfilaments at higher resolution. In \citeN{benford97a}, in which the earlier ideas, developed for the Arc
are applied to the Snake, the same battery mechanism is invoked and the development of the toroidal field
leads to pinched current loops. Using classical results relating to the relative importance of the pinch and
kink instabilities, Benford argues that the kink instability dominates when
$B_\phi \sim B_z$, explaining the kinks in the Snake. In addition, these are invoked as the site of
primary resistance and dissipation. However, the discussion of filamentation is different. Benford appeals
to a filamentary instability identified by
\citeN{molvig75a}. This requires a minimum flow speed in the pinched current defined by $v/c > 3 \times
10^{-2} (B/{\rm mG}) n_p^{-1} (2f-1)^{-1/2}$ where $f = I_r/I_z$ is the ratio of the radial to axial
currents. The minimum velocity requires electrons that are moving at much greater than the drift speed
required to maintain the magnetic field and necessarily involves the particles that are accelerated by the
dissipation. However, the minimum velocity also exceeds the Alfv\'en speed but the effect of resonant
scattering is ignored. Moreover, no reason is given as to why the current propagates over the distance
to the kinks without dissipation or for the {\em distribution} of radio flux density and spectral index in
the vicinity of the kinks. Furthermore, the ambient magnetic field
$\sim 7 \mu \rm G$ deduced by \shortciteN{gray95a} from a sound analysis of their rotation measure data
argues strongly against a pervasive magnetic field $\sim \rm mG$ in this part of the Galactic Center. To put it
another way, if one wants to argue for a mG field in this region then one has to show why the Gray et al.
rotation measure analysis is wrong.

\paragraph{Shock waves, morphologically unusual SNRs and exotica.} You are referred to \shortciteN{gray95a}
for a discussion of these.

\section{Estimates of the magnetic field in the Galactic Center}

\label{s:b-field}

As we have seen, the numerous theories for the filaments in the Galactic Center involve various assumptions
for the magnetic field -- with many theories postulating an interstellar magnetic field of the order of a
milli-Gauss. Other theories invoke a more modest magnetic field that is enhanced in some way. Hence, it is
a good idea to address this issue separately and to ask what evidence there is for a large-scale milli-Gauss
magnetic field and if there are other alternatives.

A convenient starting point for the discussion of these assumptions entails consideration of the 
{\em minimum energy} estimates of the magnetic flux density in the Galactic Center filaments.
These estimates are generally of the order of a mG and, of course, are always subject to the usual challenge:
How do you know that the filaments are in a minimum energy state? The magnetic field could be much higher or
lower. This is correct. However, such calculations do reveal that the {\em minimum pressure} of the
filaments
$\sim 10^{-8} \> \rm dyn \> cm^{-2}$ is higher than the normal value
$\sim 10^{-12} \> \rm dyn \> cm^{-2}$ for the interstellar medium so that one is presented with two choices:
(1) The Galactic Center ISM has a high total (magnetic plus thermal) pressure that confines the filaments,
consistent with the notion of a $\sim \> \rm mG$ poloidal field
\cite{serabyn94a,morris98a,chandran00a} (2) The filaments are self-confined; i.e. they are in an almost
force-free configuration with a toroidal field $B_\phi \gtapprox B_z$, where $B_z$ is the field along the
filament. The little-noted but extremely important observational deduction from Faraday rotation
measurements \shortcite{gray95a} that the  magnetic field in the vicinity of the Snake $\sim 7 \mu \rm G$ tends to
rule out a large poloidal field and therefore the second choice  becomes a real possibility. To put it another
way, if one wishes to postulate a milli-Gauss magnetic field then one has to show how the analysis of
\shortciteN{gray95a} can be plausibly altered or perhaps why it only applies to the immediate vicinity of the
Snake. 

On the other hand, \cite{serabyn94a} have, by equating the turbulent and magnetic pressures in the molecular
clouds associated with the arc, have argued for a strong magnetic field in the vicinity of Sagittarius A. (For a
refinement of these calculations, see \citeN{morris96a}.) Certainly the visual impression created by the Arc and
the Threads (see, for example, the image in \citeN{morris96a} conveys the impression of a large scale poloidal
field illuminated in places by relativistic electrons. Nevertheless, at this point we draw attention to another
feature of the observations,  namely the existence of faint, apparently {\em helical}  emission features 
projected on the linear filaments of the Arc
\cite{yusef87a,anantha91a}. Apart from the paper \cite{yusef87a} which documented the discovery of this
feature, little comment has been made on this observations. However, such helical features are the ``smoking gun''
for a force-free field since a helix is the natural force-free \ configuration and the
dynamics of force-free fields could well be the key to the entire phenomenon of filaments in the Galaxy.
Indeed, \citeN{yusef87a} offered the suggestion that reconnection, occurring between force free flux tubes of
opposite polarity could account for the energetics of the Arc. They also speculated that the northern and
southern plumes into which the Arc degenerates may be the result of weakening confining pressure. Given the
variety of solar coronal magnetic activity that takes place in a force-free environment, the scenario considered
by \citeN{yusef87a} is surely one of many and this could well be a rich source of theoretical work in the future.

If the vicinity of the Arc is indeed  a force free region, then the Arc would represent the milli-Gauss
central region of a more extended structure, possibly formed by the rotation of foot points in the Galactic
plane. (\citeN{yusef87a} had suggested that the flux tubes associated with the Arc are anchored in the
Galactic halo.) If the field in the various regions of the Galactic Center, associated with the
filaments, is indeed force-free, then this weakens the case for a pervasive large scale milli-Gauss field.

It is also apposite to mention here the estimates of magnetic fields in other regions of the Galactic Center
where the emission is thermal rather than nonthermal. For example, \citeN{killeen92a} and \citeN{plante95a}
detected fields of $2-3 \> \rm mG$ in the Circumnuclear Disc (CND). For a summary of other magnetic field
estimates in the Galactic Center see \citeN{roberts99a}. All of these observations relate to dense regions of the
Galactic Center so that it is possible that the measurements are not indicative of the tenuous gas.

If there is a large milli-Gauss field in the Galactic Center, what is its origin? Advocates of a large magnetic
appeal to \citeN{sofue87a} who outlined how magnetic field may accumulate in the Galactic Center through
diffusion from scales ~$\sim 10 kpc$. A significant poloidal field is produced because the initial field is
primordial. Such a scenario involves untested assumptions about the nature of galaxy formation and
evolution. However, no detailed models or numerical simulations have been carried out based upon it. Another
possibility is the combined dynamo mechanism  that is mediated by an activity driven outflow from the
nucleus proposed by \shortciteN{lesch89a}.

\section{A theory for the filament, the ``Snake'', involving a reconnecting coil}

\subsection{Overview}

In our own initial attempt to unravel the mystery posed by the existence of the numerous Galactic
Center filaments, we decided to concentrate on the Snake. The comprehensive observational program and the
analysis of the VLA and ATCA data reported in the paper by \shortciteN{gray95a} makes detailed
theoretical analysis and model fitting possible. It is possible that the main features of our model that we
propose may be relevant to the other filaments and possibly, the future incorporation of other processes
(in particular synchrotron cooling and distributed acceleration sites) may lead to further advances of the
entire Galactic Centre filamentary phenomenon. 

A key feature of the Snake is that its flux density is
closely associated with ``kinks'' in its structure. The flux density has local maxima at a ``major kink''
and at a ``minor kink''. Moreover the spectral index, $\alpha$, decreases
(i.e. becomes flatter) away from the major kink, highlighting the potential dynamical importance of the
kinks in explaining the dynamics of the Snake. Our model fits into the class of models involving conventional
electrodynamics and does not invoke a large poloidal magnetic field pervading the Galactic Center
region.

\begin{figure}[ht!]
\centering \leavevmode
\includegraphics[width=\textwidth]{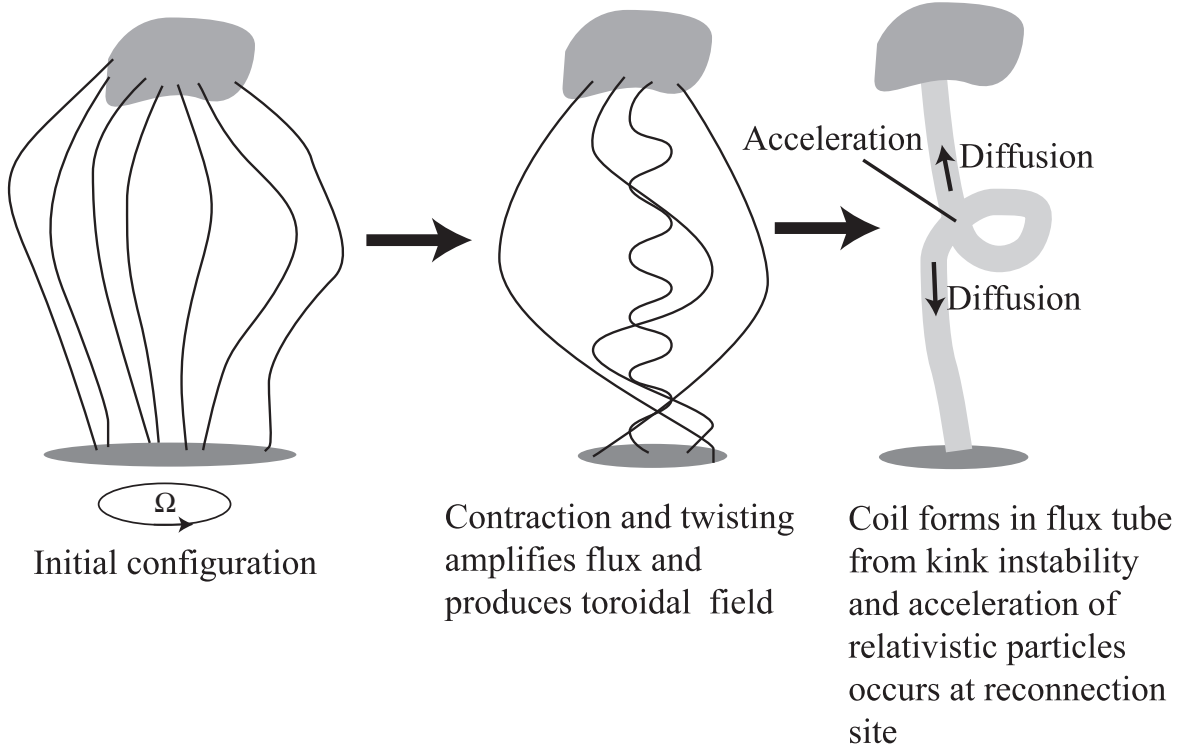}
\caption{A schematic indication of the dynamical development of a force-free flux tube that has one end
anchored in a rotating and contracting molecular cloud and the other end anchored in another part the ISM.}
\label{f:snake}
\end{figure}

It was the association of features in the radio emission with the kinks that
motivated us to think in terms of a dynamical theory for the emission that involved the classical kink
instability of a twisted magnetic flux tube. The elements of our theory are depicted schematically in
figure~\ref{f:snake} and can be summarised as follows:
\begin{enumerate}
\item The Snake is a large ($\sim 60 \> \rm pc$) magnetic flux tube, with the field lines anchored in a
rotating molecular cloud and in some other region of the ISM that may or may not rotate significantly.
Initially (i.e. before significant contraction), the field lines diverge from the molecular cloud forming
part of the network of magnetic field lines in the Galactic Center. 
\item As the cloud contracts, the field increases in the centre of the cloud and reaches a value $\sim mG$
typical of molecular cloud cores \cite{roberts99a}. 
\item At the same time, the continual twisting of the field leads to a toroidal field,
$B_\phi$ that is of the order of $B_z$ and which draws the central field lines into a thin flux tube. That
is, the flux tube becomes self-collimated and force-free. 
\item Furthermore, when $B_\phi \sim B_z$, the flux tube is unstable to the kink instability -- as in
the model of \citeN{benford97a}. This instability leads to the formation of local loops, or coils, in the
flux tube, much like the twisting of a rubber band leads to knots along its length.
\item The coils dissipate magnetic energy as a result of reconnection and possibly associated shocks.
This leads to the acceleration of relativistic particles.
\item The relativistic electrons (and, of course, the other fast particles) diffuse away from the kink
resulting in a spread of radio intensity with distance away from the kink.
\item The diffusion is energy-dependent. The highest energy particles diffuse the fastest so that the
spectral index flattens with distance from the kink.
\end{enumerate}
Our theory differs from that of \citeN{benford97a} in that the kink is produced by conventional MHD
processes, rather than by an externally induced current. It has one feature in common, namely the production
of the kink instability when $B_\phi \sim B_z$. However, the mechanism for the production of toroidal field
is quite different. Also note that the mechanisms for the production of force-free magnetic fields in the
Galactic Center is a topic that is in its infancy. Our own suggestion involving rotating molecular clouds may
be one of many and the major part of our model concerns the effect on the radio surface brightness once the
toroidal field increases to the  point where the flux tube becomes kink-unstable.

\subsection{Acceleration and diffusion of particles}

We model the Snake as a straight flux tube with spatially constant cross-sectional area and magnetic flux
density. Let $f(p,x,t)$ be the phase-space density of electrons, at time $t$ and at distance $x$ from a
reconnecting kink, located at $x=0$; let $K(p)$ be the spatial diffusion coefficient for electrons, and
$C(p)$ be the creation rate of particles per unit volume of the flux tube, per unit volume of
momentum space. We then describe the acceleration and diffusion of particles along this tube by the following
equation:
\begin{equation}
\frac {\partial f(p,x,t)}{\partial t} 
- \frac {\partial}{\partial x} \left[ K(p) \frac {\partial f(p,x,t)}{\partial x} \right]
= C(p) \delta(x)
\end{equation}
The delta function indicates that we treat the coil as a relatively small section of the entire length
of the tube. The boundary condition to this diffusion equation derived by integrating this equation
across $x=0$ is:
\begin{equation}
\frac {\partial f}{\partial x}\bigg|_{x=0} = -\frac {1}{2} \frac {C(p)}{K(p)}
\end{equation}
Throughout the period of an outburst, we assume that the injection rate, $C(p)$ is constant and that
\begin{equation}
C(p) = C_0 \, \left( \frac {p}{p_0} \right)^{-s}
\end{equation}
where $p_0 = {\rm GeV}/c$ is the fiducial value of the momentum used throughout this treatment. 
In adopting this description, the details of the injection process are not considered. A power-law with
$s \approx 4$ is relevant if strong shocks are involved.

We take the diffusion parameter also to be described by a power-law:
\begin{equation}
K(p) = K_0 \, \left( \frac {p}{p_0} \right)^\beta
\end{equation}
This dependence of the diffusion coefficient on momentum is crucial. A positive value of $\beta$ implies
that high energy electrons diffuse the fastest and thus dominate the radio emission at large distances from
the site of injection. This is our explanation for the flattening spectral index away from the kink.

In this model, we have ignored synchrotron losses of the diffusing particles. This is justified {\em a
priori} by the absence of any features in the spectrum of the Snake that could be ascribed to radiative
losses. We return to this point below in \S~\ref{s:timescales}

\subsection{Parameters and solution of the diffusion equation}

The diffusion equation is expressed in dimensionless form using a scaling length, $L$ that is also used
to define a dimensionless time variable. We have
\begin{eqnarray}
\xi &=& \frac {x}{L} \\
\tau &=& \frac {K(p)t}{L^2} = \frac {K_0 t}{L^2} \, 
\left( \frac {p}{p_0} \right)^\beta \\
g(\xi,\tau) &=& \frac{2 K_0}{ C_0 L} \, \left( \frac{p}{p_0} \right)^{s+\beta} \, f(p,x,t)
\end{eqnarray}
In these variables the diffusion equation and boundary condition take a particularly simple form which has
an analytical solution (see \citeN{bicknell01c}). The number of electrons per unit Lorentz factor is
$ N(\gamma,\xi,\tau) = N_0 \, \gamma^{-a} \, g(\xi,\tau)
$ where $N_0 = 2\pi (m_e c)^3 (C_0 L/K_0) \gamma_0^{2+a}$ and $a=s+\beta-2$. The angle-averaged synchrotron
emissivity may be estimated from $N(\gamma,\xi,\tau)$ using
 the angle-averaged single-electron synchrotron emissivity, $\bar F (y)$ defined by:
\begin{equation}
\bar F(y) = y \int_y^\infty \sqrt{1-y^2/u^2} \,  K_{5/3}(u) \> du
\end{equation} 
with $K_{5/3}(u)$ the usual modified spherical Bessel function of order $5/3$.

\subsection{Fit to the data}

\begin{figure}[ht!]
\centering \leavevmode
\includegraphics[width=\textwidth]{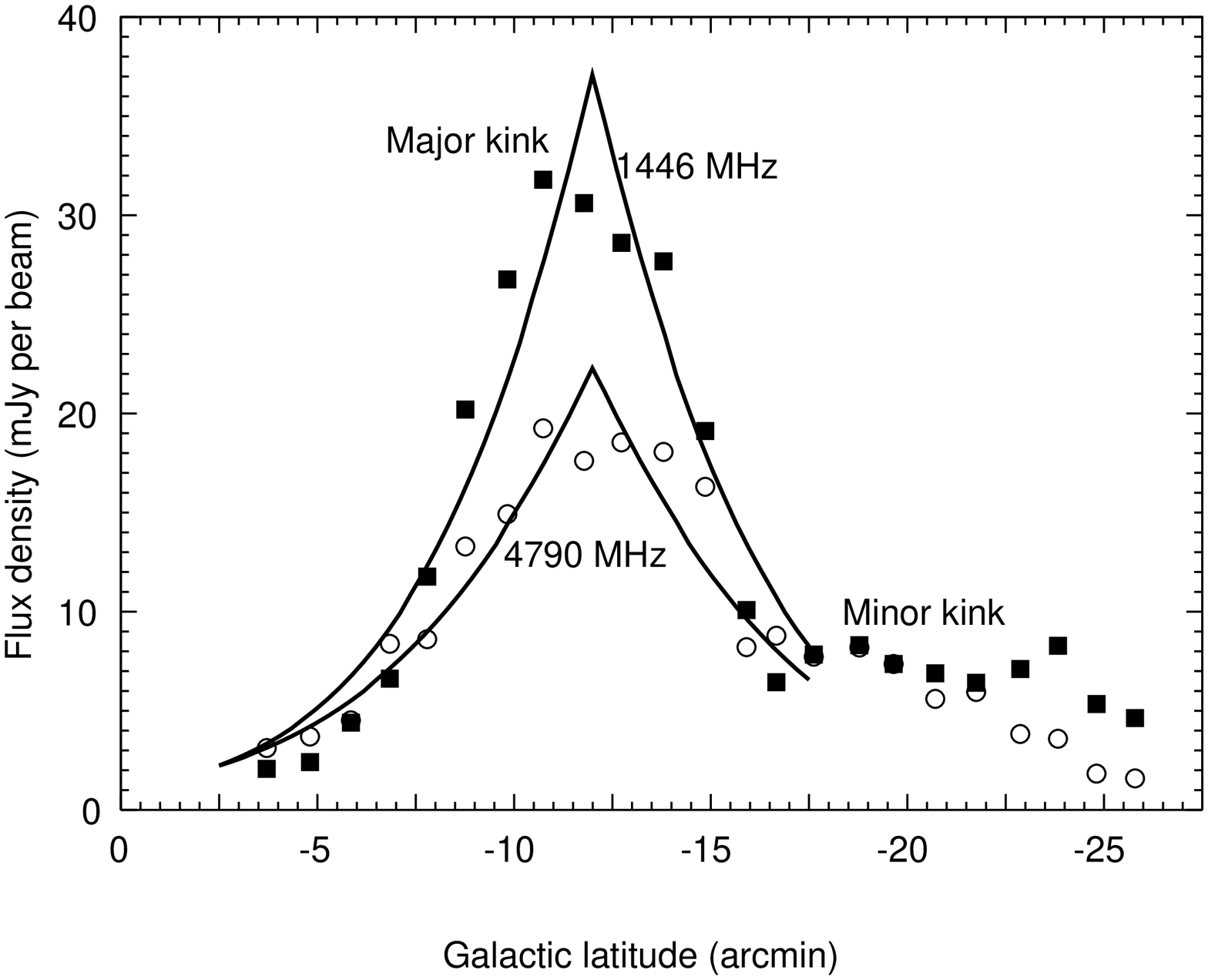}
\caption{Model fit to the data of Gray et al. (1995). The fit is restricted to the region around the
major kink in order to avoid additional parameters describing the minor kink region as well.}
\label{f:model_fit}
\end{figure}

The emissivity is used to calculate the flux density per beam along the Snake (again see \citeN{bicknell01c}
for the details), and we can then use that flux density to fit to the observational data, solving for the
parameters, $N_0, \tau_0=K_0t/L^2, B, \beta$ and $a$. We fitted that section of the data most clearly
related to the major kink in order to avoid introducing additional parameters (see
figure~\ref{f:model_fit}). Although there are 30 data points in this region, data at lower and higher
frequencies would be useful in further constraining the model and/or evaluating its predictions. 

The parameters of this fit are:
\begin{equation}
\begin{array}[]{r c l r c l r c l}
N_0 & = & 3.5 \times 10^{-5} \> \rm cm^{-3} &      B & = & 0.37 \> \rm mG & \beta & = & 0.57 \\
a & = & 2.14 \Rightarrow s = 3.57           & \tau_0 & = & {K_0 t}/{L^2} = 0.46 \nonumber
\end{array}
\end{equation}
The magnetic field is constrained by the data, albeit not very strongly, because the energy spectrum of
the emitting electrons is controlled by the diffusion and the magnetic field affects the way in which
this is reflected in the frequency domain.

There are some interesting features of the fit:
\begin{itemize}
\item The magnetic field is much higher than \shortciteN{gray95a} inferred for the interstellar medium
(ISM) in the vicinity of the Snake. This is an additional argument for the Snake flux-tube being 
force-free.
\item The value of $s$, the momentum index of the creation rate of relativistic electrons, is reasonably
close to the canonical value of 4 for shocks in a thermal medium \cite{blandford78b}
\item The value of $\beta$ is close to the model values derived for the propagation of cosmic rays in
the interstellar medium of the Galaxy. For example, \citeN{ormes83a} derived a value of
$\beta=0.8$; more recently \shortciteN{webber92a} derived a value of $\beta=0.6$. Future models may be
required to concentrate on the effects of ``minimal acceleration'' \cite{ptuskin99a} and the effects of
synchrotron cooling, the latter becoming important at higher frequencies. Nevertheless, the
agreement between our estimation of $\beta$ and its value using a similar propagation model in an
entirely different context give us confidence that the model should be taken seriously.
\item The dimensionless time $\tau_0$ does not have a special value, e.g. $\tau_0 = 10^{-3}$ or $\tau_0
= 10^6$, indicating that the model is not ``fine-tuned''.
\end{itemize}

\subsection{The energy budget}

The energy budget is an important constraint on any dynamical process. In this case, the energy
available from the annihilation of magnetic field must exceed the inferred energy in relativistic
electrons. Because of the electron spectral index, one has to invoke a high energy cutoff ($E_c$) in
order to evaluate the energy. Using the parameters of the model, the total energy of relativistic
electrons in the tube is:
\begin{eqnarray}
E_e &=& 6.3 \times 10^{44} \left( \frac {E_c}{\rm GeV} \right)^{0.43} \nonumber \\
    &=& 3.9 \times 10^{45} \> \hbox{ergs} \qquad \hbox{for} \> E_c = 10 \> \rm GeV \nonumber \\
    &=& 1.1 \times 10^{46} \> \hbox{ergs} \qquad \hbox{for} \> E_c = 100 \> \rm GeV
\end{eqnarray}
The magnetic energy stored in the coil, $E_m \approx 3.5 \times 10^{46} \> \rm ergs$. This exceeds the
energy in relativistic electrons, comfortably in the case  of a 10~GeV cutoff.

\subsection{Timescales}
\label{s:timescales}

It is important to reconcile the various timescales that are either directly or indirectly related to the
model. We did not discuss the synchrotron cooling timescale in our ApJ letter. However, this is an
important point and we discuss it in detail here.
\begin{enumerate}
\item Since the processes of twisting and kinking of the flux tube are related to the magnetic field the
relevant dynamical timescale is the Alfv\'en time, $t_A = L_{\rm tube}/v_a$, where $L_{\rm tube}$ is the
length of the tube and $v_A = B/(4 \pi \mu n m_p)^{1/2}$ is the Alfv\'en speed. Scaling to the visible
length of the Snake, \\
$t_A \approx 1.7 \times 10^5 (L_{\rm tube} / 60 {\rm pc}) (B/0.4 {\rm mG})^{-1} (n/10 \rm
cm^{-3})^{-1/2} \> \rm yrs$. (The fiducial value of $n = 10 \> cm^{-3}$ is based upon the
estimate, $n \sim 7-14 \> \rm cm^{-3}$ \shortcite{gray95a} for the ambient ISM ionised density, derived from
rotation measure variations.)
\item According to \shortciteN{amo95a} and \citeN{bazdenkov98a} a reconnecting coil disappears
``explosively'' in $1-3$ Alfv\'en times, i.e. approximately every $t_{\rm burst} \approx (1.7-5.0)\times 10^5 
(L_{\rm tube} / 60 {\rm pc}) \, (B/0.4 {\rm mG})^{-1} \, (n/10 \> cm^{-3})^{1/2} \> \rm yr$. 
\item Explosive bursts recur approximately every 5 Alfv\'en times, i.e. on a timescale $t_{\rm recur} \approx
7.5 \times 10^5 (L_{\rm tube} / 60 {\rm pc}) \, (B/0.4 {\rm mG})^{-1} \, (n/10 \> cm^{-3})^{1/2} \> \rm yr$. 
\item The time for which the electrons have been diffusing is $t_{\rm diff} = L^2 K_0^{-1} \tau_0 \approx 
3.1 \times 10^5 K_{0,26}^{-1} \> \rm yr$ where $10^{26} K_{0,26} \> \rm cm^2 \> s^{-1}$ is the value of the
diffusion parameter for GeV electrons.
\item Synchrotron cooling has not been directly incorporated into our model for the reason given above.
However, for consistency, the absence of synchrotron cooling features should be consistent with the
estimate of the magnetic field. The synchrotron cooling time is $t_{\rm syn} \approx 7.8 \times 10^4 (B/0.4 \>
{\rm mG})^{-3/2} \, (\nu/5 {\rm GHz})^{-1/2} \> \rm yrs$. 
\end{enumerate}

Validity of the model requires that the diffusive time scale be less than both the burst timescale and the
synchrotron cooling timescale. Moreover, the diffusive timescale should not be too much less than the burst
timescale, since that would imply that we are observing the Snake at a special epoch in its history.  These
constraints can be satisfied for reasonable parameters. For example, for $L_{\rm tube} \approx 60 \> \rm pc$,
$n = 10 \> \rm cm^{-3}$ and a burst lasting $3t_A$,
\begin{eqnarray}
t_{\rm diff} < t_{\rm burst} & \Rightarrow & K_{0,26} \left(\frac{B}{0.4\rm mG}\right)^{-1} > 0.62 
\nonumber \\ 
t_{\rm diff} < t_{\rm syn}   & \Rightarrow & K_{0,26} \left(\frac{B}{0.4\rm
mG}\right)^{-3/2} > 3.97
\end{eqnarray}
We could allow {\em some} variation in the magnetic field since this is not well-determined by the model.
However, even fixing $B$ at $0.4 \> \rm mG$, these constraints are satisfied for $K_{0,26} > 4$, implying
that
$t_{\rm diff} / t_{\rm burst} \ltapprox 0.16$. If only a small variation in $B$ is allowed, e.g. $B =
0.2 \> \rm mG$, then
$K_{0,26} > 1.4$ and $t_{\rm diff}/t_{\rm burst} \ltapprox 0.9$. Both of these ratios of $t_{\rm
diff}/t_{\rm burst}$ are acceptable. Note, however, that we would not wish $K_0$ to be arbitrarily high. The
diffusion timescale would then become quite short compared to the burst timescale. Therefore, we adopt
$K_{0,26} \sim 1-10$ as a working hypothesis. 

The consideration of the synchrotron timescale suggests an explanation for the difference between flat
spectrum filaments such as the Snake and the other steep spectrum filaments. We suggest that the other
filaments are older in terms of synchrotron age; their spectra may have gone through a similar
stage to that of the Snake but have since steepened. The parameters that determine this are the magnetic field
and the age of the filament so that a more general model, with allowance for synchrotron cooling,
should show that the ratio of the age to the synchrotron cooling time is larger than in the Snake. 

An appealling aspect of the above timescales is that the recurrence time of the bursts is such that it would
not be surprising to see one burst fading whilst another is bright. Thus our explanation for the minor kink
is that it represents a previous outburst that has substantially, but not completely faded.

\subsection{Resonant scattering}

This is an area of this research in which we relied very heavily on the theoretical results on the
scattering of fast particles in Don Melrose's two-volume work on Plasma Astrophysics \cite{melrose82a}. 

In order to estimate the physical time for which the present outburst has been in existence, it is necessary
to estimate the diffusion parameter, $K(p)$. We invoke a spectrum of resonant waves in which the energy per
unit wave number, $W(k)$ is given by:
\begin{equation}
W(k)  = W_0 \left( \frac {k}{k_0} \right)^{-\eta}
\end{equation}
where $k_0$ is the resonant wave number $eB/cp_0$ corresponding to $p_0 = {\rm GeV}/c$. Let the
total wave energy density be $W_m$, then using formulae given in \cite{melrose82a}, the diffusion
parameter, corresponding to $W(k)$ is:
\begin{equation}
K(p) = 1.4 \times 10^{19} \, \eta (\eta+2) 
\, \left( \frac {B}{\rm mG} \right)^{-1} \, \left( \frac {k_0 W_0} {W_m} \right)^{-1} \,
\left( \frac {p}{p_0} \right)^{2-\eta}
\end{equation}
Our value of $\beta \approx 0.6$ implies that $\eta \approx 1.4$. This is not the index of $5/3$ that one expects
from Kolmogorov turbulence. However, it is close to the index of 1.5 that one expects from
turbulence in a plasma in which the turbulent magnetic energy is comparable to the turbulent
hydromagnetic energy \cite{kraichnan65a}. (See \citeN{ruzmaikin87a}, p142~ff, for a description of the diferences
between hydrodynamic and magnetic turbulence.) 

For the ISM, cosmic ray physicists generally assume
$K_0 \sim 10^{28} \> \rm cm^2 \> s^{-1}$ (e.g.
\citeN{ptuskin99a}), implying $k_0W_0/W_m \sim 2 \times 10^{-6}$ for $B \sim 3 \mu G$. If the {\em
relative} level of turbulence (expressed by the ratio $k_0W_0/W_m$) in the Snake filament is similar to
that in the general ISM, then a magnetic field a factor of 100 times larger, gives a value of $K_0$ a
factor of 100 times smaller, as required. Why  the relative level of turbulence should behave in this
way is presumably related to the length and velocity scales of turbulence in the ISM which we shall not go into
here. However, it is clear that the dependence of
$K(p)$ upon the magnetic field is important, and in our view, is the major reason for the reduced rate of
diffusion in the Snake. As we have noted above, \shortciteN{pohl92a} derived a diffusion coefficient $\sim (2.4 -
10) \times 10^{25} \> \rm cm^2 \> s^{-1}$ for their model of the southern plume for a magnetic field 2.5 times
higher than we have estimated for the Snake. These estimates are consistent with the same relative level of
turbulence and $K_0 \propto B^{-1}$.

\section{The origin of the magnetic field in the Snake}
\label{s:origin}

In our model, the magnetic flux tube originates in the cores of molecular clouds and as we have pointed out,
this is consistent with the estimated strengths of the magnetic fields in the Galactic Center
filaments being of the order the strengths of the magnetic field in molecular cloud cores.
\citeN{uchida96a} in fact, discovered a molecular cloud -- HII region complex at the northern end of the
Snake, near the Galactic plane. The formation of stars in such clouds is complex and  may be mediated by
magnetic fields and turbulence which determine the way in which a molecular cloud or sub-regions contract to
the star-forming stage. A comprehensive description of how our model for the Snake fits into theories for
star formation and the contraction phase of molecular clouds is beyond the scope of this paper.
Nevertheless, we can, at this stage indicate how a magnetic flux tube may become twisted and indicate the
magnitude of angular velocity required.

In our ApJ Letter, we mentioned two possible regimes of magnetically dominated star formation that have
dominated research in this field for some time: Subcritical and supercritical collapse. In the former case,
the cloud is initially supported by the magnetic field and subsequently ambipolar diffusion allows the cloud to
contract (e.g.
\citeN{basu95b} and references therein). In the latter case, the cloud is not initially supported by the magnetic
field and  initially contracts, conserving angular momentum until it comes to a state of centrifugal
equilibrium. Thereafter, magnetic braking becomes important and further contraction is mediated by the radiation
of torsional Alfv\'en waves along the magnetic field (e.g. \citeN{mestel84a},\citeN{mestel99a} and references
therein). Supercritical collapse would seem to be the most favourable regime for the scenario of filament
formation that we have outlined.

However, more recently, various workers (e.g. 
\shortciteN{ballesteros99a}, \citeN{padoan99a}, \citeN{heitsch01a} and references therein) have turned to consider
more the physics of cloud collapse/contraction in a turbulent magnetised medium with the magnetostatic field
playing a less important role. Moreover, the existence of distinct clouds in the ISM is superseded by the view
that the observed clouds are essentially density inhomogeneities in a turbulent gas with a continuous velocity
field linking the various regions. In this case, the linkage between twisting of magnetic fields and star
formation is less clear although this current work tends to favour the supercritical regime but in a way that is
more complex than envisaged by \citeN{mestel84a}. Also note that if the field in the Galactic Center is as high
as a milli-Gauss, then the effects of magnetostatic fields cannot be ignored. {\em Inter alia}, this may
make the clouds more long-lived.

Most of the simulation work on contracting molecular clouds has been done with zero or small initial angular
velocity. The typical initial angular velocity possessed by a cloud forming from a smooth medium is
one-half the curl of the smooth velocity field. For this reason, \citeN{basu95a} began their simulations with an
angular velocity $\omega \sim 10^{-15} \> \rm s^{-1}$. The initial value of the angular velocity is important for
the twisting of magnetic flux tubes; if it were significantly higher than $\sim 10^{-15} \> \rm s^{-1}$ then
rapidly rotating cores could develop. The special character of the Galactic Center may be important in this
regard. Whilst the circular velocity field (see \citeN{saha96a}) may be too flat to give rise to $\omega >>
10^{-15}
\> \rm s^{-1}$, the CO longitude-velocity diagram of the inner 300~pc of the Galaxy
\citeN{brown84a} shows complex structure and a high velocity dispersion so that it is feasible that molecular
clouds within this environment would collide, generating shocks that in turn can generate a large
amount of vorticity \cite{binney74a}. 
There is one intriguing piece of evidence for rapidly rotating molecular gas in the Galactic Center. The cloud
CS1W associated with the Arc exhibits a velocity gradient of $17 \> \rm km \> s^{-1} \> pc^{-1} \approx 5.5
\times 10^{-13} \> \rm s^{-1}$ \cite{serabyn94a}. 

Given that star formation in contracting magnetised molecular clouds is a complex process that currently is not
well understood, especially in the complex flow-field of the Galactic Center, the best that we can do is
indicate the magnitude of angular velocity that is required and indicate some of the physics and rotation
velocities that are relevant to the formation of twisted magnetic flux tubes. In the following we have in mind a
scenario involving supercritical contraction and magnetic braking.

The twisting of a flux tube is linked to the radiation of angular momentum from the rotating cloud. The flux
of angular momentum, $dJ/dt$, through a flux tube of radius $R$ is given by the expression:
\begin{equation}
\frac {dJ}{dt} = 2 \pi \int_0^R \frac {r B_\phi B_z}{4 \pi} \> r \, dr
\end{equation} 
The toroidal field is  crucial for magnetic braking. In the standard case, where the field lines
are open to infinity, the angular momentum is radiated via torsional Alfv\'en waves (see \citeN{mestel99a},
p.~452). Let $\rho_0$ be the ambient density and $\Omega_0$ the angular velocity of the cloud. The toroidal
field produced in the flux tube is given by:
\begin{equation}
B_\phi = - r (4 \pi \rho_0)^{1/2} \Omega_0 \approx 
1.5 \times 10^{-6} \, \left( \frac {n}{10 \> \rm cm^{-3}} \right)^{1/2} \,
\left( \frac {\Omega_0}{\rm km \> s^{-1} \> pc^{-1}} \right) \> \rm Gauss
\end{equation}
Development of a toroidal field with a magnitude $\sim 4 \times 10^{-4} \> \rm G$ requires an angular
velocity $\sim 270 \> \rm km \> s^{-1} \> pc^{-1}$. In the context of the molecular clouds in the ISM of the
Galaxy, this is huge!

If, on the other hand, the field is anchored in another region along the flux tube, then the corresponding
solution for the toroidal field is
\begin{equation}
B_\phi = - \frac {r \Omega_0 B_0 \, t}{L_{\rm tube}}
\end{equation}
This solution for $B_\phi$ can be derived by elementary means by considering the twisting of field lines in a
tube that is slowly rotated \citeN{alfven50a}. It can also be derived from the equations for torsional Alfv\'en
waves (see \citeN{bicknell01c}). This solution is the simplest that one can consider in the current context. It
does not take into account the contraction of the cloud, nor the initial divergence of the field lines issuing
from it. Nevertheless, it is of interest to  compare this solution with that for a tube anchored at
one end only. The tube approaches its unstable configuration when $B_\phi \sim B_0$, i.e. when $t \sim L_{\rm
tube} /
\Omega_0$. If we consider $10^7 \> \rm yr$ as the maximum lifetime of a molecular cloud, then a twisted tube with
$L_{\rm tube} \sim 60 \> \rm pc$ and $r \sim 0.2 \> \rm pc$ will become unstable in $10^7 \> \rm yr$ if $\Omega
\sim 30 \> \rm km \> s^{-1} \> pc^{-1}$. This is still a high rate of rotation for a molecular cloud, but is an
order of magnitude below the initial estimate. In order for the
theory to be viable, this would have to indicative of rotation rate achieved by a cloud after contraction.
Spectroscopic observations of the Uchida et al. complex would therefore be extremely interesting. As we mentioned
above, one cloud in the Galactic Center does seem to be rotating at a rate approaching this value.

\section{Discussion}

We have summarised the observational situation and a number of theories for the curious filaments in the Galactic
Center and have discussed at some length our own theory for the Snake. In the process of this brief review and
summary of one theory, has anything been learned? The answer to this question is necessarily subjective and other
workers in this field would answer this in entirely different ways. From a purely subjective viewpoint therefore,
it seems to us that reconnection driven by some dynamical process together with electron diffusion and
synchrotron cooling are the essential ingredients for a comprehensive theory of these filaments. There is good
(circumstantial?) evidence for this: The sites of particle acceleration in the Arc are plausibly related to
reconnection brought about by the interaction of molecular clouds with the magnetic field in Sagittarius~A, the
increase in polarised emission at the crossing of two strands in G359.54+0.18 and the coincidence of peaks in
radio emission at kinks in the Snake. It also seems to us that the sometimes bifurcated, sometimes
multi-stranded, sometimes braided morphology of the filaments is indicative (or at least suggestive) of the
topological rearrangement of magnetic field lines resulting from reconnection. We have quoted some work on this
relating to twisted magnetic flux tubes and this has motivated the model we have advanced for the Snake.
Our proposal for the production of kinks and reconnection through the rotation of the
anchoring clouds stands or falls by the detection or non-detection of rapid rotation in molecular cloud/HII
region cores which intersect the filaments. However, there are other ways in which magnetic flux tubes may
interact to provide reconnection sites. Some recent work in a solar physics context involving colliding flux
tubes (albeit twisted) is that by \citeN{linton01a}. Whatever way reconnection is initiated, it seems that the
interaction between molecular clouds and filaments is strongly related to the gas dynamics of the
bar-driven accretion in the Galactic Center. Development of this theme seems to be an exciting and
productive prospect and may illuminate the processes of accretion in galactic nuclei in general.

Once electrons are accelerated at a given site they diffuse and cool, the latter mainly as a
result of synchrotron losses. We have summarised two models that take diffusion into account, and it is
interesting to note that the diffusion parameter in each case is consistent with the same level of turbulence and
the $0.4 - 1 \> \rm mG$ strength of the magnetic field. For the Snake, we have argued that
radiative losses are unimportant at the observed frequencies. However, other filaments that are older in terms of
their cooling timescales would be expected to exhibit cooling features in their spectra. Therefore, it is
unsurprising to see a variety of spectral index characteristics in the filaments. Cooling has been
successfully incorporated into the diffusive model for the southern plume (connected to the Arc) and
presumably we shall soon see diffusive plus cooling models for all of the NTFs. 

The  issue of the particle spectrum
resulting from the reconnection regions with or without associated shocks has received little attention to date.
The theory of particle acceleration in shocks is well advanced; the theory of reconnection-induced particle
acceleration less so although there has been some recent work in this area (eg. \citeN{schopper99a},
\citeN{birk01a}). A significant problem in the context of the filaments is what determines the parameters of the
electron distribution, total energy density, minimum Lorentz factor etc. 

The strength of the magnetic field in the Galactic Center permeates all of the theoretical ideas that we have
summarised. The two main contenders seem to be (1) A pervasive milli-Gauss field (2) Isolated instances of
force-free fields. The helical field structure surrounding the Arc is good evidence for the latter and
the existence of the filaments G358.85+0.47 and  G359.85+0.39 parallel to the Galactic Plane tend to argue
against the former. However, we are sure that this will continue to be a disputed point for some time and there
are counterarguments -- such as the idea that the parallel filaments mark a change in direction  of the Galactic
Center magnetic field. Extrapolating our ideas on the Snake to other filaments, we attribute the
predominance of filaments perpendicular to the plane to the lack of shear induced disruption for filaments in this
direction. (This point arose in discussion with Professor Ron Ekers following the presentation at the
Festschrift.)

\paragraph{Acknowledgements.} We are grateful to an anonymous referee for constructive comments and to Professors
Ken Freeman and James Binney for useful discussions.


\begin{thebibliography}{}

\bibitem[\protect\citeauthoryear{{Alfv\'{e}n}}{{Alfv\'{e}n}}{1950}]{alfven50a}
{Alfv\'{e}n}, H. 1950, Cosmic Electrodynamics Princeton Series in Astrophysics
  (Oxford: Clarendon Press)

\bibitem[\protect\citeauthoryear{Amo et~al.}{Amo et~al.}{1995}]{amo95a}
Amo, H., {et~al.} 1995, Phy. Rev. E,{ 51}, 3838--3841
\bibitem[\protect\citeauthoryear{{Anantharamaiah} et~al.}{{Anantharamaiah}
  et~al.}{1999}]{anantha99a}
{Anantharamaiah}, K.~R., {Lang}, C.~C., {Kassim}, N.~E., {Lazio}, T.~J.~W., \&
  {Goss}, W.~M. 1999, in ASP Conf. Ser. 186: The Central Parsecs of the Galaxy
  (San Francisco: Astronomical Society of the Pacific), 507
\bibitem[\protect\citeauthoryear{Anantharamaiah et~al.}{Anantharamaiah
  et~al.}{1991}]{anantha91a}
Anantharamaiah, K.~R., .Pedlar, A., Ekers, R.~D., \& Goss, W.~M. 1991, MNRAS,{
  249}, 262
\bibitem[\protect\citeauthoryear{{Ballesteros-Paredes}, {V{\'
  a}zquez-Semadeni}, \& {Scalo}}{{Ballesteros-Paredes}
  et~al.}{1999}]{ballesteros99a}
{Ballesteros-Paredes}, J., {V{\' a}zquez-Semadeni}, E., \& {Scalo}, J. 1999,
  ApJ,{ 515}, 286--303
\bibitem[\protect\citeauthoryear{Basu \& Mouschovias}{Basu \&
  Mouschovias}{1995a}]{basu95b}
Basu, S. \& Mouschovias, T.~Ch. 1995a, ApJ,{ 453}, 271
\bibitem[\protect\citeauthoryear{Basu \& Mouschovias}{Basu \&
  Mouschovias}{1995b}]{basu95a}
Basu, S. \& Mouschovias, T.~Ch. 1995b, ApJ,{ 452}, 386
\bibitem[\protect\citeauthoryear{{Bazdenkov} \& {Sato}}{{Bazdenkov} \&
  {Sato}}{1998}]{bazdenkov98a}
{Bazdenkov}, S. \& {Sato}, T. 1998, ApJ,{ 500}, 966--977
\bibitem[\protect\citeauthoryear{Benford}{Benford}{1988}]{benford88a}
Benford, G. 1988, ApJ,{ 333}, 735
\bibitem[\protect\citeauthoryear{Benford}{Benford}{1997}]{benford97a}
Benford, G. 1997, ApJ,{ 333}, 735
\bibitem[\protect\citeauthoryear{Bicknell \& Li}{Bicknell \&
  Li}{2001}]{bicknell01c}
Bicknell, G.~V. \& Li, J. 2001, ApJL,{ 548}, L69--L72
\bibitem[\protect\citeauthoryear{Bicknell \& Melrose}{Bicknell \&
  Melrose}{1982}]{bicknell82a}
Bicknell, G.~V. \& Melrose, D.~B. 1982, ApJ,{ 262}, 511
\bibitem[\protect\citeauthoryear{{Binney}}{{Binney}}{1974}]{binney74a}
{Binney}, J. 1974, MNRAS,{ 168}, 73--92
\bibitem[\protect\citeauthoryear{Birk, Crusius-Waetzel, \& Lesch}{Birk
  et~al.}{2001}]{birk01a}
Birk, G.~T., Crusius-Waetzel, A.~R., \& Lesch, H. 2001, astro-ph/106565
\bibitem[\protect\citeauthoryear{{Blandford} \& {Ostriker}}{{Blandford} \&
  {Ostriker}}{1978}]{blandford78b}
{Blandford}, R.~D. \& {Ostriker}, J.~P. 1978, ApJL,{ 221}, L29--L32
\bibitem[\protect\citeauthoryear{{Brown} \& {Liszt}}{{Brown} \&
  {Liszt}}{1984}]{brown84a}
{Brown}, R.~L. \& {Liszt}, H.~S. 1984, ARAA,{ 22}, 223--265
\bibitem[\protect\citeauthoryear{{Chandran}, {Cowley}, \& {Morris}}{{Chandran}
  et~al.}{2000}]{chandran00a}
{Chandran}, B.~D.~G., {Cowley}, S.~C., \& {Morris}, M. 2000, ApJ,{ 528},
  723--733
\bibitem[\protect\citeauthoryear{{Gray} et~al.}{{Gray} et~al.}{1991}]{gray91a}
{Gray}, A.~D., {Cram}, L.~E., {Ekers}, R.~D., \& {Goss}, W.~M. 1991, Nature,{
  353}, 237--239
\bibitem[\protect\citeauthoryear{Gray et~al.}{Gray et~al.}{1995}]{gray95a}
Gray, A.~D., Nicholls, J., Ekers, R.~D., \& Cram, L.~E. 1995, ApJ,{ 448},
  164--178
\bibitem[\protect\citeauthoryear{{Heitsch}, {Mac Low}, \& {Klessen}}{{Heitsch}
  et~al.}{2001}]{heitsch01a}
{Heitsch}, F., {Mac Low}, M., \& {Klessen}, R.~S. 2001, ApJ,{ 547}, 280--291
\bibitem[\protect\citeauthoryear{Heyvaerts, Norman, \& Pudritz}{Heyvaerts
  et~al.}{1988}]{heyvaerts88a}
Heyvaerts, J., Norman, C., \& Pudritz, R.~E. 1988, ApJ,{ 330}, 718
\bibitem[\protect\citeauthoryear{Killeen, Lo, \& Crutcher}{Killeen
  et~al.}{1992}]{killeen92a}
Killeen, N.~E.~B., Lo, K.~Y., \& Crutcher, R. 1992, ApJ,{ 385}, 585
\bibitem[\protect\citeauthoryear{Kraichnan}{Kraichnan}{1965}]{kraichnan65a}
Kraichnan, R.~H. 1965, Phys. Fluids,{ 8}, 1385--1387
\bibitem[\protect\citeauthoryear{{Lang} et~al.}{{Lang} et~al.}{1999}]{lang99c}
{Lang}, C.~C., {Anantharamaiah}, K.~R., {Kassim}, N.~E., \& {Lazio}, T.~J.~W.
  1999, ApJL,{ 521}, L41--L44
\bibitem[\protect\citeauthoryear{{Lang}, {Morris}, \& {Echevarria}}{{Lang}
  et~al.}{1999}]{lang99b}
{Lang}, C.~C., {Morris}, M., \& {Echevarria}, L. 1999, ApJ,{ 526}, 727--743
\bibitem[\protect\citeauthoryear{{LaRosa} et~al.}{{LaRosa}
  et~al.}{2000}]{larosa00a}
{LaRosa}, T.~N., {Kassim}, N.~E., {Lazio}, T.~J.~W., \& {Hyman}, S.~D. 2000,
  AJ,{ 119}, 207--240
\bibitem[\protect\citeauthoryear{LaRosa, Lazio, \& Kassim}{LaRosa
  et~al.}{2001}]{larosa01a}
LaRosa, T.~N., Lazio, T.~J.~W., \& Kassim, N. 2001, astro-ph/0108360
\bibitem[\protect\citeauthoryear{{Lesch} et~al.}{{Lesch}
  et~al.}{1989}]{lesch89a}
{Lesch}, H., {Crusius}, A., {Schlickeiser}, R., \& {Wielebinski}, R. 1989,
  A\&A,{ 217}, 99--107
\bibitem[\protect\citeauthoryear{{Li} \& {Melrose}}{{Li} \&
  {Melrose}}{1994}]{li94a}
{Li}, J. \& {Melrose}, D.~B. 1994, MNRAS,{ 270}, 687
\bibitem[\protect\citeauthoryear{{Linton}, {Dahlburg}, \& {Antiochos}}{{Linton}
  et~al.}{2001}]{linton01a}
{Linton}, M.~G., {Dahlburg}, R.~B., \& {Antiochos}, S.~K. 2001, ApJ,{ 553},
  905--921
\bibitem[\protect\citeauthoryear{{Liszt} \& {Spiker}}{{Liszt} \&
  {Spiker}}{1995}]{liszt95a}
{Liszt}, H.~S. \& {Spiker}, R.~W. 1995, ApJS,{ 98}, 259
\bibitem[\protect\citeauthoryear{Melrose}{Melrose}{1982}]{melrose82a}
Melrose, D.~B. 1982, Plasma Astrophysics (London: Gordon \& Breach)
\bibitem[\protect\citeauthoryear{Mestel}{Mestel}{1999}]{mestel99a}
Mestel, L 1999, Stellar Magnetism The International Series of Monographs on
  Physics (Oxford: Clarendon Press)
\bibitem[\protect\citeauthoryear{{Mestel} \& {Paris}}{{Mestel} \&
  {Paris}}{1984}]{mestel84a}
{Mestel}, L. \& {Paris}, R.B. 1984, A\&A,{ 136}, 98
\bibitem[\protect\citeauthoryear{Molvig}{Molvig}{1975}]{molvig75a}
Molvig, K. 1975, Phys. Rev. Lett.,{ 35}, 22
\bibitem[\protect\citeauthoryear{{Morris}}{{Morris}}{1998}]{morris98a}
{Morris}, M. 1998, in IAU Symp. 184: The Central Regions of the Galaxy and
  Galaxies, Volume 184 331
\bibitem[\protect\citeauthoryear{Morris \& Serabyn}{Morris \&
  Serabyn}{1996}]{morris96a}
Morris, M. \& Serabyn, E. 1996, ARAA,{ 34}, 645--701
\bibitem[\protect\citeauthoryear{Morris \& Yusef-Zadeh}{Morris \&
  Yusef-Zadeh}{1985}]{morris85a}
Morris, M. \& Yusef-Zadeh, F. 1985, AJ,{ 90}, 2511--2513
\bibitem[\protect\citeauthoryear{Morris \& Yusef-Zadeh}{Morris \&
  Yusef-Zadeh}{1989}]{morris89a}
Morris, M. \& Yusef-Zadeh, F. 1989, ApJ,{ 343}, 703--712
\bibitem[\protect\citeauthoryear{Nicholls \& {Le~Strange}}{Nicholls \&
  {Le~Strange}}{1995}]{nicholls95a}
Nicholls, J. \& {Le~Strange}, E.~T. 1995, ApJ,{ 443}, 638
\bibitem[\protect\citeauthoryear{{Ormes} \& {Protheroe}}{{Ormes} \&
  {Protheroe}}{1983}]{ormes83a}
{Ormes}, J.~F. \& {Protheroe}, R.~J. 1983, ApJ,{ 272}, 756--764
\bibitem[\protect\citeauthoryear{{Padoan} \& {Nordlund}}{{Padoan} \&
  {Nordlund}}{1999}]{padoan99a}
{Padoan}, P. \& {Nordlund}, A., 1999, ApJ,{ 526}, 279--294
\bibitem[\protect\citeauthoryear{Plante \& anf R.~M.~Crutcher}{Plante \& anf
  R.~M.~Crutcher}{1999}]{plante95a}
Plante, R.~L. \& R.~M.~Crutcher, anf K.~Y.~Lo 1999, in The central parsecs of
  the Galaxy, ed. H.~Falcke, A.~Cotera, W.~J. Duschl, F.~Melia, \& M.~J. Rieke,
  Volume 186 of ASP Conference Series (San Francisco: Astronomical Society of
  the Pacific), 483--487
\bibitem[\protect\citeauthoryear{{Pohl}, {Reich}, \& {Schlickeiser}}{{Pohl}
  et~al.}{1992}]{pohl92a}
{Pohl}, M., {Reich}, W., \& {Schlickeiser}, R. 1992, A\&A,{ 262}, 441--454
\bibitem[\protect\citeauthoryear{Ptuskin et~al.}{Ptuskin
  et~al.}{1999}]{ptuskin99a}
Ptuskin, V.~S., Lukasiak, A., Jones, F.~C., \& Webber, W.~R. 1999, in 26th
  International Cosmic Ray Conference, ed. D.~Kieda, M.~Salamon, \& B.~Dingus,
  Volume~4 291--294
\bibitem[\protect\citeauthoryear{Reich}{Reich}{1990}]{reich90a}
Reich, W. 1990, in IAU Symposium 140: Galactic and Intergalactic Magnetic
  Fields, ed. R.~Beck, P.~Kronberg, \& R.~Wielebinski (Dordrecht: Kluwer), 369
\bibitem[\protect\citeauthoryear{Roberts}{Roberts}{1999}]{roberts99a}
Roberts, D.~A. 1999, in ASP Conf. Ser. 186: The Central Parsecs of the Galaxy
  483
\bibitem[\protect\citeauthoryear{{Rosner} \& {Bodo}}{{Rosner} \&
  {Bodo}}{1996}]{rosner96a}
{Rosner}, R. \& {Bodo}, G. 1996, October), ApJL,{ 470}, L49
\bibitem[\protect\citeauthoryear{Ruzmaikin, Shukurov, \& Sokoloff}{Ruzmaikin
  et~al.}{1987}]{ruzmaikin87a}
Ruzmaikin, A.~A., Shukurov, A.~M., \& Sokoloff, D.~D. 1987, Magnetic Fields of
  Galaxies (Dordrecht: Kluwer)
\bibitem[\protect\citeauthoryear{Saha, Bicknell, \& McGregor}{Saha
  et~al.}{1996}]{saha96a}
Saha, P., Bicknell, G.~V., \& McGregor, P.~J. 1996, ApJ,{ 467}, 636
\bibitem[\protect\citeauthoryear{Schopper, Birk, \& Lesch}{Schopper
  et~al.}{1999}]{schopper99a}
Schopper, R., Birk, G.~T., \& Lesch, H. 1999, Physics of Plasmas,{ 6},
  4318--4327
\bibitem[\protect\citeauthoryear{{Serabyn} \& {Morris}}{{Serabyn} \&
  {Morris}}{1994}]{serabyn94a}
{Serabyn}, E. \& {Morris}, M. 1994, ApJL,{ 424}, L91--L94
\bibitem[\protect\citeauthoryear{{Shore} \& {Larosa}}{{Shore} \&
  {Larosa}}{1999}]{shore99a}
{Shore}, S.~N. \& {Larosa}, T.~N. 1999, ApJ,{ 521}, 587--590
\bibitem[\protect\citeauthoryear{{Sofue} \& {Fujimoto}}{{Sofue} \&
  {Fujimoto}}{1987}]{sofue87a}
{Sofue}, Y. \& {Fujimoto}, M. 1987, Pub. Astr. Soc. Japan,{ 39}, 843--848
\bibitem[\protect\citeauthoryear{Uchida et~al.}{Uchida
  et~al.}{1996}]{uchida96a}
Uchida, K.~I., Morris, M., Serabyn, E., \& G\"usten, R. 1996, ApJ,{ 462},
  768--776
\bibitem[\protect\citeauthoryear{{Webber}, {Lee}, \& {Gupta}}{{Webber}
  et~al.}{1992}]{webber92a}
{Webber}, W.~R., {Lee}, M.~A., \& {Gupta}, M. 1992, ApJ,{ 390}, 96--104
\bibitem[\protect\citeauthoryear{Yusef-Zadeh \& Morris}{Yusef-Zadeh \&
  Morris}{1987}]{yusef87a}
Yusef-Zadeh, F. \& Morris, M. 1987, ApJ,{ 322}, 721
\bibitem[\protect\citeauthoryear{Yusef-Zadeh, Morris, \& Chance}{Yusef-Zadeh
  et~al.}{1984}]{yusef84a}
Yusef-Zadeh, F., Morris, M., \& Chance, S. 1984, Nature,{ 310}, 557--561
\bibitem[\protect\citeauthoryear{{Yusef-Zadeh}, {Wardle}, \&
  {Parastaran}}{{Yusef-Zadeh} et~al.}{1997}]{yusef97a}
{Yusef-Zadeh}, F., {Wardle}, M., \& {Parastaran}, P. 1997, ApJL,{ 475}, L119
\end{thebibliography}
\end{document}